\documentclass[]{article}
\usepackage{graphics,amsmath,fancyhdr,amssymb}
\usepackage[centerlast,sc,small]{subfigure}

\title{Ground--$\gamma$ band mixing and evolution of collectivity in even-even
neutron-rich nuclei with ${40 <}$Z$ < 50$}

\begin{document}

\maketitle

\begin{centering}

\author{S.~Lalkovski$^1$ and N.~Minkov$^2$} \\

\vspace{0.5cm}

$^1${\it Department of Physics, University of Sofia, 1164 Sofia, Bulgaria}\\

$^2${\it Institute of Nuclear Research and Nuclear Energy, 72 Tzarigrad Road, 1784 Sofia, Bulgaria}\\

\end{centering}

\vspace{0.5cm}

\abstract{
We propose an extended band mixing formalism capable of describing
the ground-$\gamma$ band interaction in a wide range
of collective spectra beyond the regions of well deformed nuclei.
On this basis we explain the staggering effects observed in the
$\gamma$ bands of Mo, Ru and Pd nuclei providing a consistent
interpretation of new experimental data in the neutron
rich region. As a result the systematic
behavior of the odd-even staggering effect and some general
characteristics of the spectrum such as the mutual disposition of
the bands, the interaction strength and the band
structures is explained as the manifestation of respective changes
in collective dynamics of the system.

}

\pagestyle{fancy}
\renewcommand{\headrulewidth}{1.3pt}
\renewcommand{\baselinestretch}{1.3}
\fancyhead[]{}
\fancyhead[LO,LE]{S. Lalkovski and N. Minkov, {\it Ground--$\gamma$ band mixing ...}} 
\fancyhead[RO,LE]{\thepage}
\fancyfoot[]{}

\section{Introduction}

The low lying excited states of even-even nuclei are usually
described in a geometrical approach as the levels corresponding to
harmonic vibrations, rotations of deformed shapes or unstable
shape rotations \cite{BM75}. These three geometrical models have
been associated with the symmetry limits of the Interacting Boson
Model (IBM) \cite{IA87}, in which the low-lying excited states are
classified according to the irreducible representations of three
chains of subgroups of the group U(6), labelled as U(5), SU(3) and
O(6). These symmetries are considered as the stable limits of
collectivity in nuclear structure. However, most of nuclei have a
transitional behavior taking place in regions between the above
mentioned symmetries. Recently it has been suggested that
some additional symmetries as E(5) \cite{Ia00,CZ00} and
X(5) \cite{Ia01,CZ01},
might take place between U(5) and O(6) and between
U(5) and SU(3) respectively, emphasizing the need for extensive
study of the ways in which nuclear collective properties deviate
from the above ``standard'' symmetries.

The odd-even staggering effect observed in the $\gamma$ bands is
among the most sensitive phenomena carrying information about the
symmetry changes. It is well pronounced in nuclear
regions characterized by U(5) and O(6) and
relatively weaker in nuclei near the SU(3) region.  In the latter
case the staggering effect can be reproduced through the
$\gamma$-$\beta$ band mixing interaction. In the U(5) and O(6)
regions the gsb and $\gamma$ bands are strongly coupled to each other
and the effect can be explained on the basis of the
gsb-$\gamma$ band mixing interaction.

A detailed theoretical study
of the ground-$\gamma$ band coupling mechanism has been
implemented in the framework of the Vector Boson Model with SU(3)
dynamical symmetry \cite{Mi97,Mi00}. It suggests a
relevant model interpretation of the ground-$\gamma$ band mixing
interaction and explains the related odd-even staggering effects
in terms of the SU(3) multiplets inherent for the underlying
algebraic scheme.

In the framework of the collective algebraic models the
presence or the lack of the staggering effect as well as its
magnitude and form give a specific information for the appearance
or absence of particular symmetry characteristics of the spectrum.
\begin{figure}[h]\centering
\rotatebox{-90}{\scalebox{0.36}[0.36]{\includegraphics{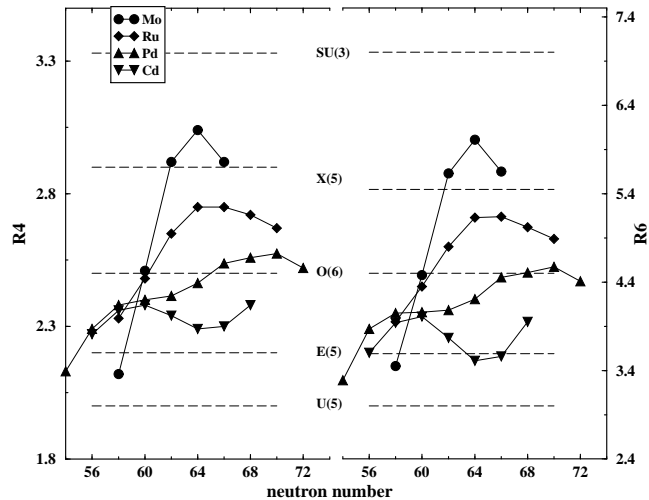}}}
\caption[]{The excitation energy ratios $R_4$ and $R_6$ (see text)
in several Mo, Ru, Pd and Cd isotopes are plotted as function of
neutron number.} \label{R}
\end{figure}

From experimental point of view, essential progress in the
collective nuclear structure study has been made through the use
of a new generation of multidetector $\gamma$-ray arrays, such as
Eurogam \cite{NBF94}, Euroball \cite{Si97} and Gammasphere
\cite{Lee90}, which provides  opportunity to investigate the
prompt $\gamma$-rays emitted from fission fragments. In the last
decade this technique has been applied in a number of experiments
on both spontaneous and induced fission. In such a way excited
states in neutron-rich nuclei $^{100-108}$Mo
\cite{SL01,Ho91,PRC1191,EPJ175}, $^{104-114}$Ru \cite{EPJ177,PLB136} and
$^{108-118}$Pd \cite{EPJ,EPJ151,JPG2253} have been
populated.

The new data provide a rather detailed test for the way of the
changes in the symmetries mentioned above. This is illustrated in
fig.~\ref{R} where the gsb excitation energy ratios $R_4 =
E(4)/E(2)$ and $R_6 = E(6)/E(2)$ are plotted as functions of the
neutron number. For example one sees that the addition or
subtraction of few neutrons in Mo isotopic chain leads to rapid
changes from near U(5) ($^{100}$Mo) to near SU(3) symmetry
($^{106}$Mo). Recently it has been suggested that  X(5) phase
transition between the above symmetries occurs in $^{104}$Mo
\cite{Bi02}. These changes are less pronounced in the Ru and Pd
isotopic chains. The collectivity in Ru isotopes gradually
increases from U(5) towards SU(3) symmetry limit, while in Pd
nuclei it develops towards O(6) \cite{Ki96}. In the framework of
the shape-phase transition concept the nuclei $^{104}$Ru and
$^{108}$Pd are proposed as possible candidates for E(5) critical
behavior \cite{FAA01,ZL02}. The study of the shape evolution of
$^{102-108}$Ru in terms of the potential energy surfaces shows an
increase in the deformation parameter $\beta$ with the neutron
number \cite{EPJ177}. The staggering effect in the $\gamma$-bands
of these nuclei is explained as the manifestation of $\gamma$
softness. The evolution of the shape in dependence on the angular
momentum was studied in ref.~\cite{Re03} for the nuclei in the
region $40<Z<50$. It suggests a change in collectivity within the
yrast line from vibrational to rotational.

Furthermore, the new experimental data provide extended higher spin
structure of collective bands, especially in gsb and $\gamma$ bands,
which deserves an adequate interpretation in
the general framework of the ``changing collectivity'' concept.
Thus, the systematic analysis of extended data suggests that
the states of the ground band and the
$\gamma$-band interact in a way similar to what is observed in
rotational SU(3) nuclei in the framework of the Vector Boson Model
(VBM) \cite{Mi97}. This circumstance suggests a possible scenario
of a transition between rotational and vibrational collective
spectra in which the same ground-$\gamma$ band coupling mechanism
persists while the collective band structure is changed. That is
way the ground and $\gamma$-band states can be a subject
of interest in the examination of the basic changes in nuclear
collectivity. We only remark that while in the rotational SU(3)
region the term $\gamma$-band is well defined in association with
a rotation built on a $\gamma$-vibrational state, in nuclei
towards transitional and vibrational regions this geometrical
meaning is not clear anymore.

Based on the above, in the present paper we propose that it will
be reasonable to apply an extended model in which the mixing
scheme inherent for the SU(3) dynamical symmetry of VBM is
modified so as to take into account the changes in the level
spacing characteristics. We suggest that it will be capable to
reproduce the changes in the band structure between rotational,
transitional and vibrational modes. In this framework, our purpose
is to implement a detailed analysis and interpretation of the
ground-$\gamma$ band interaction in a wide range of collective
nuclei beyond the exact symmetry limits of collectivity.

We consider the nuclear region $40<Z<50$, providing a model
interpretation of the experimental ground and $\gamma$-band levels
and their interaction together with the respective odd-even
staggering phenomena observed there. As it will be shown below,
the obtained results reveal the systematic behavior of the
inter-band interaction and provide a relevant systematics of the
$R_4$ excitation energy ratios in dependence on a specific factor
of collectivity.

In sec.~\ref{gsbg} we present a formalism for a unique
description of the gsb - $\gamma$-band interaction. In sec.
\ref{nrd} model descriptions for the energy levels and the
corresponding staggering patterns of nuclei in the region $40 < Z
< 50$ are presented. A detailed analysis of the systematic changes
in nuclear collectivity is presented also there. In sec. \ref{On}
some algebraic aspects of the interplay between rotation and
vibration modes and the possible ways of its involvement in the
study are discussed. In sec. \ref{co} concluding remarks are
given.

\section{Ground-$\gamma$ - band mixing formalism and odd-even staggering}
\label{gsbg}

The odd-even staggering effect represents a relative displacement
of the even angular momentum levels of the $\gamma$-band with
respect to the odd levels. It has been explained in the framework
of the Vector Boson Model trough the interaction of the even
$\gamma$-band levels with their counterparts in the gsb
\cite{Mi00}.

The basic assumption of the VBM is that the low lying collective
states of deformed even-even nuclei can be reproduced through the
use of vector bosons, whose creation operators are O(3) vectors.
The angular momentum operator ${\bf L}$ as well as the quadrupole
operator ${\bf Q}$ are constructed by these vector bosons. The VBM
Hamiltonian is constructed as linear combination of three O(3)
scalars from the enveloping algebra of SU(3): ${\bf L}^2$, ${\bf
L.Q.L}$ and ${\bf A}^\dagger {\bf A}$ \cite{Mi97}, where the
second and the third terms are third and fourth order effective
interactions, reducing the SU(3) symmetry to the rotational group
O(3). In the framework of VBM the gsb and $\gamma$ - bands belong
to the same split SU(3) multiplet labelled by the quantum numbers
($\lambda, \mu$). In this framework the model provides a relevant
way to study the interaction between these two bands. In the
simplest case of multiplets of the type ($\lambda ,2$) the even
counterparts of both bands are mixed trough a secular equation of
the form
\begin{equation}
\begin{array}{|cc|}
V_{11}-E & V_{12}\\
V_{21}   & V_{22}-E\\
\end{array}
=0 \ ,
\label{secular}
\end{equation}
where $V_{ij}$ $(i,j=1,2)$ are the matrix elements of the model
Hamiltonian in the used SU(3) basis states, known as the basis of
Bargmann-Moshinsky \cite{BM61,MPSW}. For a given angular momentum
there are two solutions (corresponding to the each band)
containing terms of the form $L(L+1)$ \cite{Mi00}. This formalism
has been applied in rare-earth and actinide regions giving a
successful description of the odd-even staggering effect in the
$\gamma$-bands.
\begin{figure}[h]\centering
\rotatebox{-90}{\scalebox{0.35}[0.35]{\includegraphics{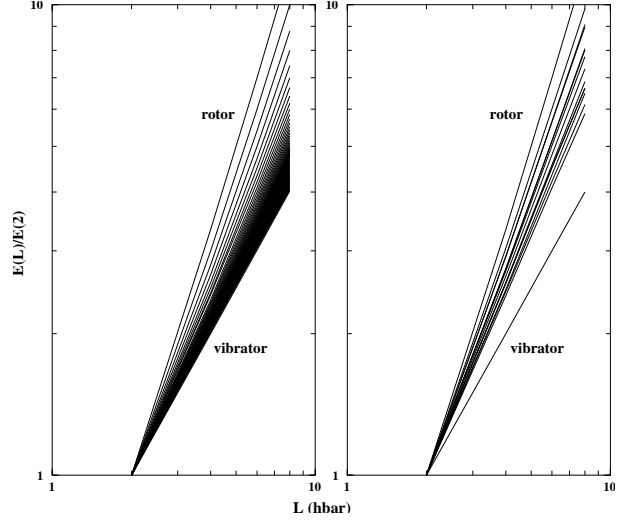}}}
\caption[]{The excitation energy ratio $E(L)/E(2)$ is given vs
angular momentum for the: (a) energy term $E(L)=L(L+n)$ with $n=1,
2, 3, ...$; (b) experimental gsb levels of Mo, Ru, Pd nuclei.}
\label{E-I}
\end{figure}

However, Pd and Ru nuclei are far from the SU(3) symmetry regions,
so that the original VBM formalism cannot be applied directly.
Therefore, in order to develop a unique framework for description
of the staggering effect on the basis of the gsb-$\gamma$ - band
interaction we have to modify properly the rotational term
$L(L+1)$ with a generalized expression capable to generate
an additional linear
dependence of the energy levels $\sim L$. In such a way the
changes of the spectrum from SU(3) to U(5) direction can be taken
into account. For this reason we introduce a global
parameter of collectivity, which is denoted here as `$n$' and
provides a modified angular momentum dependence of the energy in
the form $L(L+n)$. For $n \sim 1$ it gives the level spacing of a
good rotor while for large values of $n$ it gives a level spacing
close to that of a vibrator as it is demonstrated in
fig.~\ref{E-I}. For intermediate $n$-values this term gives a
transitional spectrum. As it will be demonstrated below the most
deformed nucleus in $Z\sim 50$ region, $^{106}$Mo, is described by
$L(L+3)$, while the nucleus $^{100}$Mo nearest to U(5) limit by
$L(L+26)$.

We suppose that the model Hamiltonian can be modified so that a
linear $L$-dependence be generated by a diagonal term $V'$
providing the following form of the secular eq.~(\ref{secular}):
\begin{equation}
\begin{array}{|cc|}
 \widetilde{V}_{11} -E & V_{12}\\
V_{21}   & \widetilde{V}_{22}-E\\
\end{array}
=0 \ ,
\label{ss}
\end{equation}
where
\begin{equation}
\widetilde{V}_{ii}=V_{ii}+V'\, ,  \qquad  i=1,2\, .
\end{equation}
Here the off-diagonal elements remain the same, following the
assumption that the same gsb-$\gamma$ band coupling mechanism is
conserved, while the collective band structure is changed.

In this way eq.~(\ref{ss}) provides a modified expression for the gsb
and $\gamma$-band level energies:
\begin{equation}
\begin{split}
E_{g,\gamma} &=\frac{1}{2}[ V_{11}+V_{22}+2V' \\
     & +(-1)^{\nu_{g,\gamma}} \sqrt{(V_{11}-V_{22})^2+4V_{12}V_{21}}\, ] \\
\end{split}
\label{equ4}
\end{equation}
($\nu_g=1$ for gsb; $\nu_{\gamma}=0$ for $\gamma$) in which the
energy splitting term (the square root) inherent for VBM remains
the same. Then the original VBM energy expressions (eqs (14) and
(15) of ref. \cite{Mi00}) can be extended in the following form:
\begin{multline}
E_{g,\gamma}(L)=AL(L+n)+2B{1+(-1)^{\nu_{g,\gamma}} \over 2}\\
+(-1)^{\nu_{g,\gamma}} B[\sqrt{1+aL(L+1)+bL^2(L+1)^2}\\
-(-1)^{\nu_{g,\gamma}} CL(L+n)-1]{1+(-1)^L \over 2} \ ,
\label{su3ene}
\end{multline}
where
\begin{equation}
\nu_{g,\gamma} =
\begin{cases}
0& \gamma -\mathrm{band}\\
1& \mathrm{ground\ state\ band} \ .
\end{cases}
\label{ene}
\end{equation}

In the framework of VBM the quantities $A$, $B$, $C$, $a$
and $b$ are functions of the strengths in the effective
interaction in the model Hamiltonian \cite{Mi00}
and on the SU(3) quantum numbers ($\lambda, \mu$). In the
present modification the  additional
term $V'$, generating the energy dependence of the type $L(L+n)$,
is not derived in explicit form. The reason is that any particular
assumption for its construction at this stage of the study will
impose a limitation on the generality of analysis and
interpretation of experimental data.
Possible ways to determine this term and some algebraic aspects
of its physical meaning are discussed in sec.~\ref{On}.
For the same reason we do not impose the relation between
the quantities in eq.~(\ref{su3ene}) and the quantum numbers of
the group SU(3). Hence, we consider them as model parameters
that have to be determined directly on the basis of
experimental data.

We have to remark that a global parameter of collectivity similar
to ``n'' has been used in a Sp(4,R) classification scheme with
respect to the low-lying ground band states of even-even nuclei
\cite{Dr95}. It has been shown that such an approach successfully
reproduces the changes in the ground band structure from
rotational to vibrational collective regions.

In eq.~(\ref{su3ene}) the even levels of the gsb interact with
their $\gamma$ band counterparts through the square root term.
This term together with the term containing the parameters $C$
causes the relative shift between the odd and even states in the
$\gamma$-band, i.e. it generates an odd-even staggering effect.

As a relevant characteristics of the staggering effect we consider
the following three-point formula \cite{Bo88}
\begin{equation}
\delta E(L) = E(L) - {{(L+1)E(L-1)+LE(L+1)}\over {2L+1}} \ ,
\label{stgform}
\end{equation}
where $E(L)$  denotes the energy of the level with angular
momentum $L$. Obviously, for the simple rigid rotor energy
$L(L+1)$ one has $\delta E=0$. Thus any energy dependence with
$\delta E \not = 0$ will correspond to respective deviation from
the regular rotor behavior of the system. As it will be seen in
the next section, this characteristics carries detailed
information for the evolution of collectivity in different nuclear
regions.

{\scriptsize
\begin{table}
\caption{\label{tab:table1}Theoretical end experimental gsb and
$\gamma$ energy levels in keV for Mo nuclei. The fitting
parameters $A$ and $B$ are given in keV, while the parameters $C$,
$a$, $b$ and $n$  are dimensionless. The RMS factors $\sigma_E$,
eq. (\protect \ref{rms}), are also given in keV.}
\centering
\begin{tabular}{ccccccc}
\hline\noalign{\smallskip} Nucl./param., Ref. & L & E$^{th}_{gsb}$
& E$^{exp}_{gsb}$
& E$^{th}_{\gamma}$  & E$^{exp}_{\gamma}$\\
\noalign{\smallskip}\hline\noalign{\smallskip}
$^{100}$Mo $n=26$ \cite{SL01} &  2 &   534 &  536    & 1103  &1064\\
$\sigma_E = 31.65$ &  3 &       &     & 1540  &1607\\
A= 11.5468 &  4 &  1136 &  1136     & 1759  &1771\\
B= 267.8468 &  5 &        &     & 2325  &2288\\
C= 0.0064 &  6 &  1828 &  1847  &   &\\
a= 0.0229 &    &          &     &   &\\
b= -0.0003 &  8 &  2637 &  2626     &   &\\
\hline
$^{102}$Mo $n=7$ \cite{Ho91} &2&    314&    296& 876 &   848\\
$\sigma_E = 19.62$  &3&        &       & 1210&   1246\\
A= 22.5967     &4&    755&    743& 1387&   1398\\
B= 266.0827  &5&       &       & 1888&   1870\\
C= 0.0162  &6&   1322&   1328&&\\
a= 0.0186   &7&       &       &&\\
b= 0.0001   &8&   2015&   2019&&\\
\hline
$^{104}$Mo $n =5$ \cite{PRC1191}
                & 2&    223 &   192 & 835 &    812\\
$\sigma_E = 15.29$& 3&        &       &1009 &   1028\\
A= 17.3603      & 4&   573  &  561  &1216 &   1215\\
B= 296.3152     & 5&        &       &1461 &   1475\\
C= 0.0025       & 6&   1065 &  1080 & 1720&   1724\\
a= 0.0122       & 7&        &       & 2051&   2037\\
b= -0.0002      & 8&   1723 &  1722 & 2325&   2326\\
\hline
$^{106}$Mo $n= 3$ \cite{EPJ175}
 & 2&    188&    172& 722&    710\\
$\sigma_E =8.07$  & 3&         &       & 876 &   885\\
A=     19.2189  & 4&    528&    523& 1068&   1068\\
B=    264.9240  & 5&       &       & 1299&   1307\\
C=       0.0006  & 6&   1026&   1033& 1561&   1563\\
a=      0.0031  & 7&       &       & 1875&   1868\\
b=    -0.0001  & 8&   1689&   1688& 2194&   2194\\
\hline
$^{108}$Mo  $n= 3$ \cite{PRC1191}
  & 2&    202&    193& 611 &   586\\
$\sigma_E = 11.77$  & 3&       &       & 776 &   783\\
A= 20.1844 & 4&    564&    564& 970 &   978\\
B= 206.3506  & 5&        &       & 1220&   1232\\
C= 0.0008 & 6&   1084&   1091& 1491&   1508\\
a= -0.0028  & 7&         &       & 1826&   1817\\
b= 0.0001  & 8&   1755&   1752& 2180&   2170\\
\noalign{\smallskip}\hline
\end{tabular}\end{table}
}

\section{Numerical results and discussion}\label{nrd}

In order to reproduce the ground and $\gamma$-band energy levels
we adjust the model parameters $A$, $B$, $C$, $a$ and $b$ with
respect to the corresponding experimental data by using a $\chi^2$
minimization procedure at fixed value of the parameter $n$. Since
$n$ is a discrete quantity, we apply the procedure consequently in
a wide range $n=1-100$. As a result we determine the $n$-value
which provides the smallest root mean square (RMS) deviation
\begin{equation}
\sigma _E =\sqrt{{1 \over
N_B}\sum_{L,\nu}[E_\nu^{th}(L)-E_\nu^{exp}(L)]^2} \ , \label{rms}
\end{equation}
where $N_B$ is the number of the levels used in the fitting
procedure and $\nu =g, \gamma$ for the gsb and $\gamma$-band
respectively. We restrict our calculations up to the back-bending
region. That is why the two bands are considered up to angular
momentum $L=8 \hbar$.

{\scriptsize
\begin{table}
\caption{\label{tab:table2} The same as in table~\ref{tab:table1}
but for Ru isotopes.}\centering
\begin{tabular}{ccccccc}
\hline\noalign{\smallskip} Nucl./param., Ref. & L   &
E$^{th}_{gsb}$ &  E$^{exp}_{gsb}$
& E$^{th}_{\gamma}$ & E$^{exp}_{\gamma}$ \\
\noalign{\smallskip}\hline\noalign{\smallskip} $^{104}$Ru $n=15$
\cite{EPJ177}
  &2&    405&    358& 920 &   893 &\\
$\sigma_E =23.16$   &3&        &       & 1216&   1242&\\
A=     14.1863   &4&    898&    888& 1508&   1502&\\
B=    224.9008   &5&       &       & 1868&   1872&\\
C=       0.0058   &6&   1521&   1556& 2173&   2196&\\
a=      0.0560   &7&       &       & 2634&   2623&\\
b=    -0.0007   &8&   2331&   2320& 2855&   2847&\\
\hline $^{106}$Ru $n=7$ \cite{EPJ177}
   &2 &   302&    270&  834 &  792\\
$\sigma_E =22.81$  & 3 &       &       & 1067 &  1091\\
A=     18.2663  & 4 &   732&    715& 1295 &  1307\\
B=    259.4464  & 5 &      &       & 1615 &  1642\\
C=       0.0043  & 6 &  1288&   1297& 1905 &  1908\\
a=      0.0083  & 7 &      &       & 2309 &  2285\\
b=     0.0000  & 8 &  1970&   1975&      &\\
\hline $^{108}$Ru $n=6$ \cite{PLB136}
   &2&    269&    242& 741 &   708\\
$\sigma_E=17.83$   &3&         &       & 954 &   974\\
A=     18.6971   &4&    671&    665& 1180&   1183\\
B=    224.6123   &5&       &       & 1478&   1497\\
C=       0.0052   &6&   1223&   1241& 1749&   1762\\
a=      0.0189   &7&       &       & 2151&   2133\\
b=    -0.0002   &8&   1948&   1942& 2426&   2421\\
\hline $^{110}$Ru $n=5 $ \cite{PLB136}
  & 2&    265&    241& 646 &   613\\
$\sigma_E =17.01$  & 3&           &       & 849 &   860\\
A=     19.7289  & 4&    677&    663& 1074&   1085\\
B=    187.6927  & 5&           &       & 1362&   1376\\
C=       0.0033  & 6&   1236&   1239& 1662&   1685\\
a=      0.0050  & 7&           &       & 2033&   2021\\
b=     0.0000  & 8&   1938&   1944& 2412&   2398\\
\hline
$^{112}$Ru $n=5$ \cite{PLB136}
                  &2&    251&    237& 554    &524\\
$\sigma_E = 15.52$   &3&       &       & 736&    748\\
A=     18.6310   &4&    641&    645& 975  &  981\\
B=    144.3896   &5&       &       & 1220 &  1236\\
C=       0.0013   &6&   1173&   1190& 1550&   1571\\
a=      0.0171   &7&       &       & 1854 &  1841\\
b=     0.0000   &8&   1850&   1840& 2275  & 2264\\
\noalign{\smallskip}\hline
\end{tabular}\end{table}
}

The procedure has been applied to the nuclei $^{100-108}$Mo,
$^{104-112}$Ru and for $^{108-116}$Pd. As a result the respective
gsb and $\gamma$-band energy levels have been reproduced quite
accurately. This is demonstrated in tables~\ref{tab:table1},
\ref{tab:table2}, and \ref{tab:table3}, where the obtained
theoretical descriptions are compared with the experimental data.
The resulting model parameters and RMS factors $\sigma_E$ are also
given there. The $\sigma_E$-values indicate the good quality of
model description, considering the relatively large collective
energy values in the nuclear region under study. For example the
RMS factors generally do not exceed 5-7{\%} of the energy of the
corresponding lowest (gsb) $2^+$ states.

As it is seen from tables~\ref{tab:table1}--\ref{tab:table3}, the
parameters values vary smoothly within and between the isotope
groups of the considered nuclear region giving additional
indication for the model quality. For example the parameter $B$
vary in the limits 130-230 keV for Pd isotopes, 140-260 keV for
Ru, and 200-300 keV in Mo-isotopes. Together with the values of
the higher order parameters $a$ and $b$ ($|a|\sim 10^{-2}$,
$|b|\sim 10^{-4}$), the obtained $B$-values determine [according
to eq.~(\ref{su3ene})] the mutual disposition of the $\gamma$ band
and the gsb and provide the increase in the band shift from Pd to
Mo isotopes, which will be discussed in details below. Further,
the values of the parameter $A$ varying in the limits 11-22 keV,
together with the product $BC$ (appearing as a correction to $A$)
determine the inertial characteristics of the two bands. The
behavior of the parameter of collectivity $n$ is also discussed in
details below, so here we just remark that it is a stable model
characteristics. As a discrete integer quantity it is determined
with an error not larger than one as well as within a well
determined ``RMS minimum'' regions in the range $n=1-100$.

Since the agreement between the theory and the experiment is good,
the analysis and the conclusions for the considered nuclear
characteristics made hereafter are valid on the basis of both
theoretical and experimental data. So, unless it is specified (by
``exp'' or ``th'') any comment on a nuclear quantity will refer
equally to the respective experimental and theoretical values.

\begin{figure}[h]\centering
\rotatebox{0}{\scalebox{0.55}[0.55]{\includegraphics{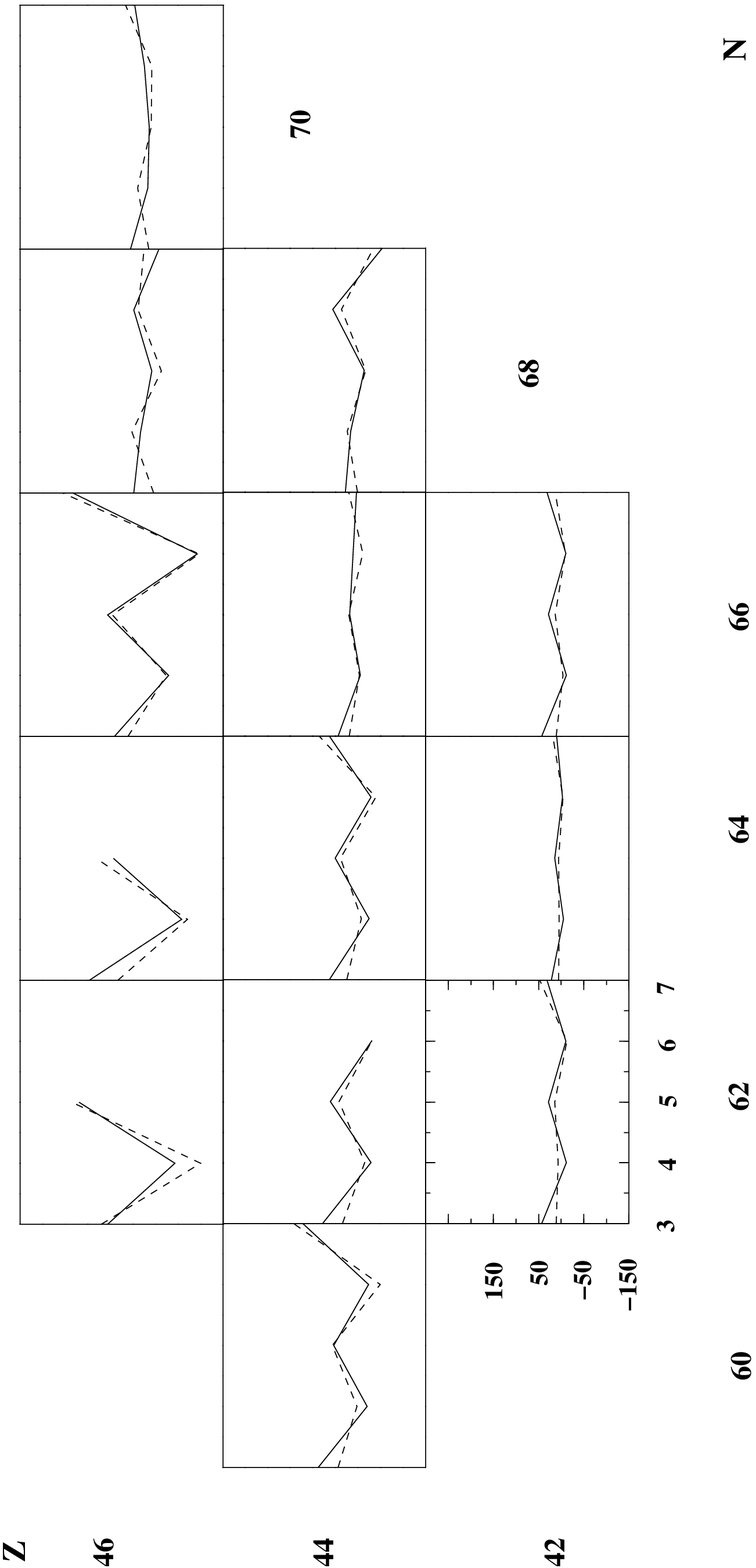}}}
\caption[]{Theoretical (dashed lines) and experimental (solid
lines) odd-even staggering plots (in keV) [see
eq.~(\protect\ref{stgform})] for $^{104-108}$Mo, $^{104-112}$Ru
and $^{108-116}$Pd nuclei.} \label{stagg-MRP}
\end{figure}
In fig.~\ref{stagg-MRP} we present the theoretical and
experimental staggering plots for Mo, Ru and Pd isotopes obtained
by applying eq.~(\ref{stgform}) to the respective theoretical and
experimental energy levels of the $\gamma$ bands. It is seen that
the staggering effect in these nuclei is described successfully,
with the respective phase and amplitude characteristics of the
staggering patterns being reproduced accurately.

The following observations and comments can be made. In
$^{108-112}$Pd isotopes the staggering effect is well pronounced,
in spite of the small number of data available. In $^{114,116}$Pd
the staggering amplitude is strongly suppressed, and moreover in
$^{116}$Pd the oscillations are almost reduced. Ru isotopes with
N=60, 62 and 64 exhibit staggering patterns with almost equal
amplitudes, weaker than the amplitudes observed in $^{108-112}$Pd
isotopes. In $^{110}$Ru the staggering effect is again strongly
suppressed. In $^{104-108}$Mo isotopes the observed staggering
amplitudes are generally smaller compared to Pd and Ru nuclei. On
the above basis we deduce that for the considered nuclei the
increasing neutron number and decreasing number of protons lead to
a systematic suppression of the odd-even staggering effect in the
$\gamma$-bands. In such a way a region of a relatively better
formed rotation structure in these bands is outlined.

{\scriptsize
\begin{table}
\caption{\label{tab:table3} The same as in table~\ref{tab:table1}
but for Pd nuclei.}\centering
\begin{tabular}{ccccccc}
\hline\noalign{\smallskip} Nucl./param., Ref. & L   &
E$^{th}_{gsb}$ &
E$^{exp}_{gsb}$ & E$^{th}_{\gamma}$ & E$^{exp}_{\gamma}$\\
\noalign{\smallskip}\hline\noalign{\smallskip}
$^{108}$Pd $n=16$ \cite{EPJ}
   &2&    475&    434& 959 &   931\\
$\sigma_E =28.46$    &3&       &       & 1341&   1335\\
A=     15.3942   &4&   1051&   1048& 1570&   1625\\
B=    231.7060   &5&       &       & 2080&   2083\\
C=       0.0082   &6&   1741&   1772& 2283&   2259\\
a=      0.0162   &7&       &       && \\
b=    -0.0002   &8&   2562&   2549&& \\
\hline
$^{110}$Pd $n=12$ \cite{EPJ}
   &2&    403&    374& 861&    814\\
$\sigma_E =28.37$   &3&       &       & 1163&   1212\\
A=     15.5428   &4&    922&    921& 1369&   1398\\
B=    231.8628   &5&       &       & 1785&   1759\\
C=       0.0054   &6&   1557&   1574& 1992&   1987\\
a=     -0.0041   &7&       &       &&\\
b=     0.0000   &8&   2302&   2296&&\\
\hline $^{112}$Pd $n=13$ \cite{EPJ151}
   &2 &   391 &   348& 781 &    736\\
$\sigma_E = 27.08$   &3 &       &      & 1084&   1096\\
A=     15.4425   &4 &   884 &   883& 1339&   1362\\
B=    171.2245   &5 &       &      & 1732&   1759\\
C=       0.0094   &6 &  1512 &  1550& 1983&   2002\\
a=      0.0540   &7 &       &      & 2504&   2483\\
b=    -0.0008   &8 &  2336 &  2318& 2652&   2638\\
\hline $^{114}$Pd $n=15$ \cite{JPG2253}
&2&    389&    333& 736&    695\\
$\sigma_E =28.68$      &3&       &       & 989&   1012\\
A=     13.2847   &4&    864&    852& 1314&   1321\\
B=    135.9780   &5&       &       & 1600&   1631\\
C=       0.0055   &6&   1462&   1501& 1970&   1984\\
a=      0.1118   &7&       &       & 2318&   2290\\
b=    -0.0013   &8&   2227&   2216& 2660&   2655\\
\hline
$^{116}$Pd $n=13$ \cite{JPG2253}
   &2&   394&   341 & 762 &   738\\
$\sigma_E=22.75$   &3&        &       & 1042&   1067\\
A=     15.9262   &4&   889&   878 & 1379&   1374\\
B=    138.6619   &5&      &     & 1711&   1719\\
C=       0.0093  &6&   1526&  1559& 2088&   2101\\
a=      0.1356   &7&       &        & 2507&   2493\\
b=    -0.0015   &8&   2352&  2344& 2843&   2840\\
\noalign{\smallskip}\hline
\end{tabular}\end{table}
}

For comparison, in rare-earth and actinide nuclei a gradual
decrease of the staggering effect is observed towards the
mid-shell regions \cite{Mi00}. It is explained with the respective
better pronounced (less perturbed) rotational band structures
there. In the framework of VBM this is related with a decrease in
the gsb-$\gamma$ band interaction strength. Here, on the same
basis we can interpret the suppressed staggering effect as the
result of a decreasing interaction strength between the gsb and
$\gamma$-band.

{\scriptsize
\begin{table}
\caption{\label{tab:table5} Experimental and theoretical
excitation ratios and band disposition factors (see the text) in
Mo, Ru and Pd nuclei. The mass number $A$ is also given.}\centering
\begin{tabular}{cccccccccc}
\hline\noalign{\smallskip} Nucl. &A  & $\Delta E_2^{exp}$ &
$\Delta E_2^{th}$ & $R_3^{exp}$ &
$R_3^{th}$ & $R_4^{exp}$ & $R_4^{th}$\\
\noalign{\smallskip}\hline\noalign{\smallskip}
Mo
&100 & 0.99 & 1.07 & 1.30 & 1.50 & 2.12 & 2.13 \\
&102 & 1.86 & 1.79 & 1.38 & 1.53 & 2.51 & 2.40 \\
&104 & 3.23 & 2.74 & 1.87 & 2.19 & 2.92 & 2.57 \\
&106 & 3.13 & 2.84 & 2.05 & 2.25 & 3.04 & 2.81 \\
&108 & 2.04 & 2.02 & 1.99 & 2.18 & 2.92 & 2.79 \\
\hline
Ru
&104 & 1.49 & 1.27 & 1.74  & 1.99  & 2.48 & 2.22 \\
&106 & 1.93 & 1.76 & 1.72  & 1.98  & 2.65 & 2.42 \\
&108 & 1.93 & 1.75 & 1.79  & 2.06  & 2.75 & 2.49 \\
&110 & 1.54 & 1.44 & 1.91  & 2.11  & 2.75 & 2.55 \\
&112 & 1.21 & 1.21 & 2.04  & 2.31  & 2.72 & 2.55 \\
\hline
Pd
&108 & 1.15  &  1.02 &  1.72 & 1.50 & 2.41 & 2.21 \\
&110 & 1.18  &  1.14 &  1.47 & 1.68 & 2.46 & 2.29 \\
&112 & 1.11  &  1.00 &  1.74 & 1.84 & 2.54 & 2.26 \\
&114 & 1.09  &  0.89 &  1.97 & 2.28 & 2.56 & 2.22 \\
&116 & 1.16  &  0.93 &  1.93 & 2.20 & 2.57 & 2.26 \\
 \noalign{\smallskip}\hline
\end{tabular}\end{table}
}

Now we will consider another important characteristics of the
interacting ground and $\gamma$ bands
\begin{equation}
\Delta E_2 = {E_\gamma(2)-E_g(2) \over E_g(2)} \ , \label{El}
\end{equation}
which describes their mutual disposition. In terms of VBM,
eq.~(\ref{El}) corresponds to the splitting of a SU(3) state with
even angular momentum $L=2$ into the levels of the gsb and the
$\gamma$-band, i.e. it characterizes the splitting of the SU(3)
multiplet. It has been shown that in well deformed nuclei the
larger splitting is associated with the weaker band mixing
interaction which is the case observed in the mid-shell regions of
rare-earth and actinide nuclei  \cite{Mi99}. Moreover, it has been
demonstrated that the extremely large magnitude of the splitting
could be related to a situation of completely separated
(noninteracting) bands, known from group theoretical point of view
as an SU(3) contraction process. Similar analysis can be done for
the nuclear region under study. However, now we should have in
mind that away from the exact SU(3) region the term
ground-$\gamma$ multiplet has not the same clear group theoretical
meaning, so that the term ``splitting'' should simply refer to the
mutual disposition of the bands.

So, the splitting ratio $\Delta E_2$ carries an information about
the systematic changes in the mutual band disposition and
respective band-mixing interaction in dependence on the place of
the nucleus in a given region. It has been shown \cite{Mi97} that
in the rare-earth region this value is within the limits $7<
\Delta E_2 < 18$, while in the actinides it is between $13< \Delta
E_2 < 25$.

The experimental and theoretical values of $\Delta E_2$, obtained
for nuclei in the region $40 < Z < 50$ are given in
table~\ref{tab:table5}. We see that in $^{104,106,108}$Mo $\Delta
E^{exp}_2$ varies between 2.0 and 3.2. For $^{104-112}$Ru it is
between 1.2 and 2.0, while in Pd nuclei $\Delta E^{exp}_2$ is
about 1.1-1.2. We remark that in the considered nuclear region the
relative displacement of the ground and the $\gamma$ band is
essentially smaller compared with the rare-earth and actinide
regions, which is naturally a prerequisite for their essentially
stronger mutual perturbation. Here $\Delta E_2$ overally decreases
from Mo to Pd nuclei as an indication for the respective
increasing of the band mixing interaction.

\begin{figure}[h]\centering
\rotatebox{-90}{\scalebox{0.35}[0.35]{\includegraphics{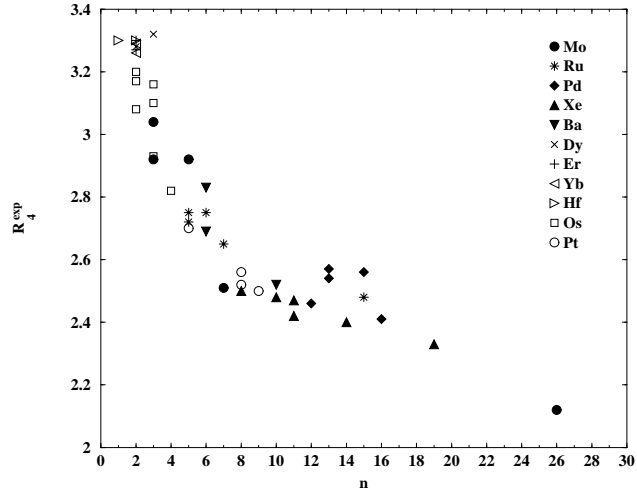}}}
\caption[]{The experimental excitation energy ratio $R^{exp}_4$ is
given vs. the parameter $n$ for $^{100-108}$Mo, $^{104-112}$Ru and
$^{108-116}$Pd nuclei. Also, results for the nuclei $^{116-126}$Xe
\cite{Sakai,To94,Sch99}, $^{124,128,130}$Ba
\cite{Pi90,Sakai,Su85}, $^{162,164}$Dy \cite{Sakai},
$^{164,168}$Er \cite{Sakai}, $^{168,172}$Yb \cite{Sakai},
$^{176,178}$Hf \cite{Sakai}, $^{180,192}$Os \cite{Sakai} and
$^{182,186-190}$Pt \cite{Sakai,He90} are included for better
statistics.} \label{R4-n}
\end{figure}

In fig.~\ref{R4-n} we present the systematic relation between the
experimental excitation ratio
$R^{exp}_4=E_{g}^{exp}(4)/E_{g}^{exp}(2)$ and the obtained values
of the model parameter of collectivity $n$. Additional results
providing better statistics and pointing the applicability of the
used model scheme in the neighboring region of Xe and Ba as well
as in the well deformed rare earth nuclei Er, Yb, Hf, Os, and Pt
are also given there. So, fig.~\ref{R4-n} demonstrates the way in
which the collective properties of nuclei under study deviate from
the SU(3) symmetry. We see that starting from the SU(3) rotational
region with values near $3.33$, $R^{exp}_4$ rapidly decreases with
the increase of $n$ until reaching the region near
$R^{exp}_4=2.5$. After that, in the region $n=8-16$ we observe an
overall saturation towards values typical for the ground state
bands with a structure between transitional and vibrational.

In the regions $n\sim 20$, $n > 20$ there is an indication (in
$^{116}$Xe and $^{100}$Mo) for further gradual decrease in
$R^{exp}_4$ with the increase of $n$. However, these two points
are not enough to draw a definite conclusion about this region of
$n$. Moreover, the $\gamma$ band in $^{100}$Mo (see table 1) is
observed up to $L=5$, so that future data on the higher states
might refine the position of this nucleus in the scheme. On the
other hand the presence of a gap between $^{100}$Mo and the other
nuclei in fig.~\ref{R4-n} is not unexpected. This is seen from the
$R^{exp}_4$ plots in fig. 1. We remark that for the nuclei
$^{100-106}$Pd and $^{102}$Ru appearing at this figure with
$R^{exp}_4$ between 2.15-2.48 there is no available points in fig.
4, due to the insufficient data on the $\gamma$- bands. In this
respect $^{100-106}$Pd and $^{102}$Ru are placed between the main
group of considered nuclei and $^{100}$Mo for which
$R^{exp}_4=2.12$. Thus one may expect that future experimental
data on the above group of nuclei can fill the gap in fig. 4.

\begin{figure}[h]\centering
{\scalebox{0.52}[0.52]{\includegraphics{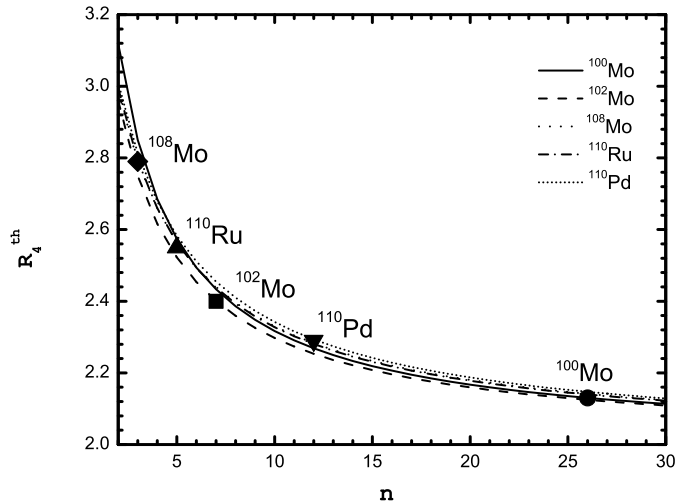}}}
\caption[]{Analytical behavior of the theoretical ratio
$R^{th}_4=E_{g}^{th}(4)/E_{g}^{th}(2)$ as a function of the
parameter $n$ (family of curves) for the sets of parameters $A$,
$B$, $C$, $a$ and $b$ of the nuclei $^{100,103,108}$Mo, $^{110}$Ru
and $^{110}$Pd (see tables~\ref{tab:table1}--\ref{tab:table3}).
The $R^{th}$-values obtained for the given nuclei (see
table~\ref{tab:table5}) are indicated on the respective curves.}
\label{R4Th}
\end{figure}

Nevertheless, we are now capable to analyze the general systematic
behavior of $R^{exp}_4$ in fig. 4 through the analytical behavior
of the theoretical ratio
\begin{eqnarray}
R^{th}_4=E_{g}(4)/E_{g}(2) \nonumber
\end{eqnarray}
obtained from eq.~(\ref{su3ene}) as a function of the parameter
$n$. This is illustrated in fig. 5, where $R^{th}_4$ is plotted
for the sets of parameters ($A$, $B$, $C$, $a$ and $b$) of the
nuclei $^{100,103,108}$Mo, $^{110}$Ru and $^{110}$Pd. The
$R^{th}_4$-values fixed for the given nuclei (through the fitting
procedure) are indicated on the respective curves. First, we
remark that the observed family of curves outlines the way in the
$R^{th}_4$--$n$ plane in which nuclear collectivity evaluates. The
rapid decrease in $R_4$ (starting from about 3.2) near the
beginning of the $n$-axis is obvious. Further we see that in the
region $n=5-10$, the slope of the curves decreases, so that after
$n=10-15$ the $R^{th}_4$ values decrease slowly. The further
behavior of the curves shows that for very large $n$'s the ratio
$R^{th}_4$ saturates towards the vibration value 2.00.

Looking again on fig.4 we see that the overall behavior of the
experimental $R_4$ ratios can be interpreted in the same way.
However, now we can indicate more specifically that in the region
$n=12-16$ the $R^{exp}_4$  correlation with $n$ looks a bit
perturbed compared to other regions. Also in this region the slope
of (overall) decrease in $R^{exp}_4$ with $n$ is not well
pronounced.

This observation can be explained as follows. While $R_4$ is a
characteristics of the lowest ground state band structure the
global parameter $n$ carries additional information about the
structure of the ground band and the gsb-$\gamma$ bands
interaction as well. Having this in mind we can conclude that the
well pronounced correlation between $R_4$ and $n$ observed in the
region of near-SU(3) nuclei may be considered as the result of a
smooth evolution of collectivity in the two bands. Although  in
the case of near-SU(3) nuclei the gsb and $\gamma$-band
essentially interact, their mutual perturbation is still not
strong enough and conserves the individual characteristics of the
two bands. As a result $R_4$ remains a good characteristics of
collectivity not only for the gsb structure, but also for the
entire low-lying spectrum in the given nucleus. In the region $n
\geq 10$ (see fig.~\ref{R4-n}) the situation is different.
According to our band-mixing scheme, there the mutual perturbation
of the two bands should be quite strong. Therefore, the
information that the gsb characteristics $R_4$ could carry for the
$\gamma$-band structure will be essentially limited, so it will
not characterize anymore the complicated ground - $\gamma$ band
configuration, and the overall collective properties of the given
nucleus. For example it could happen that the gsb is characterized
by near rotational $R_4$ value with the $\gamma$-band carrying
some characteristics of a far from rotational structure. On the
other side the global parameter $n$ is continuously capable of
taking into account the common collective characteristics of the
gsb and $\gamma$ band structures (and their interaction) in the
different nuclei.

In the above aspect we remark that the physical meaning of $n$
considerably differs from the gsb systematic characteristic
$\omega$ used in the Sp(4,R) classification scheme \cite{Dr95}.
This is indicated by the larger $n$-values compared to the
respective values of $\omega$ reported in \cite{Dr95}. (Compare
tables 1-3 of this work with tables 3-4 in \cite{Dr95}) We note
that far from the vibration U(5) nuclei this difference is not too
large, for example in $^{106}$Mo one has $n=3$ and $\omega=2$, in
$^{110-112}$Ru $n=5$ and $\omega=3$. In nuclei towards the U(5)
region the difference is quite large, for example in $^{104}$Ru
$n=15$ while $\omega=7$, in $^{100}$Mo $n=26$ while $\omega=12$.
This observation clearly shows that the presence of the
$\gamma$-band (and its interaction) in our collective scheme is of
crucial importance for the evolution of collectivity in nuclei
beyond rotational regions, especially in the nuclei towards U(5).

Now, we find it worth to add one more comment on the region
$n=12-16$ for $R^{exp}_4$ in fig. 4. The almost reduced decrease
in $R^{exp}_4$ in this region and its possible faster drop
expected for $n>20$ may be referred to an essential rearrangement
in collectivity between $R_4=2.2-2.5$ including the development of
the O(6) $\gamma$-softness (starting from $R^{exp}_4=2.5$) with
the possible further appearance of E(5) phase transition near
$R^{exp}_4=2.2$ ($n>20$). However we should remark that further
detailed study in this direction exceeds the current framework of
the formalism used and moreover the insufficient data in the
region $n>20$ do not allow us to draw more detailed conclusion.

In table 4 we give the theoretical and experimental values for the
energy ratio
\begin{equation}
R_3 ={E_{\gamma}(4)-E_{\gamma}(2)\over
E_{\gamma}(3)-E_{\gamma}(2)} \ , \label{r3}
\end{equation}
used in \cite{Sakai} as a characteristics of the $\gamma$-band
collective structure. In the limit of a pure rotational
band-structure, it takes the values 2.33 while in the vibration
limit, where the states $3^+$ and $4^+$ merge into a degenerate
vibration state, $R_3$ takes the value 1.00.

We see that in the considered region both $R^{exp}_3$ and
$R^{th}_3$ exhibit a trend of decrease from near rotational to
near vibrational nuclei (see table~\ref{tab:table5}), i.e.
decrease with increasing $n$. For example in $^{106}$Mo
$R^{exp}_3=2.05$ while in $^{100}$Mo it drops to 1.30. However, we
should remark that the $R_3$ ratio is not generally stable
characteristics of collectivity compared to $R_4$, especially in
the regions of the strong band-mixing interaction. The reason is
that as seen from eq.~(\ref{r3}), the $R_3$ ratio involves both
even ($2^+$, $4^+$), and odd ($3^+$) states. Hence, it is strongly
affected by the odd-even shift, i.e. by the staggering effect.

\begin{figure}[h]\centering
\rotatebox{-90}{\scalebox{0.35}[0.35]{\includegraphics{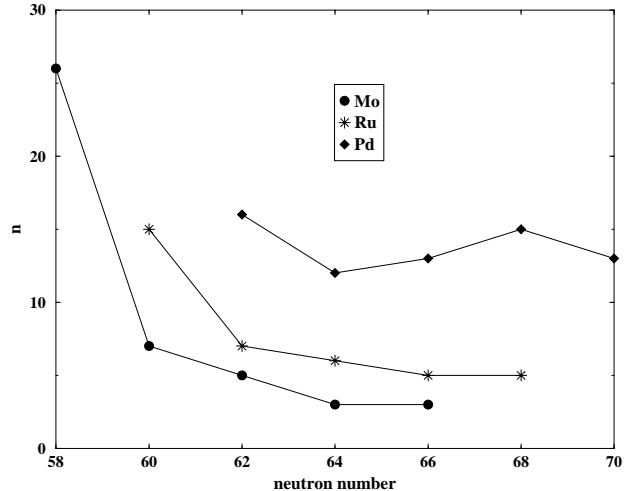}}}
\caption[]{The model parameter $n$ vs. neutron number.} \label{nN}
\end{figure}

The so far obtained results allow us to reveal the specific
signatures of the changing nuclear collectivity in terms of the
respective symmetries. Thus the local saturation of $R^{exp}_4$
with respect to $n$, observed in fig.~\ref{R4-n} (the region
$R^{exp}_4=2.5$) can be interpreted as a manifestation of the
$\gamma$-softness structure inherent for the O(6) symmetry. Also,
the gradual decrease in the staggering effect observed in
fig.~\ref{stagg-MRP} can be interpreted as the general result of
the change from SU(3) to O(6) symmetry. The nuclei with near
U(5)-SU(3) [or X(5)] transition collective structure
($R^{exp}_4\sim 2.9$) are characterized by $\Delta E^{exp}_2$
values between 2.0 and 3.2, indicating a weak interband
interaction strength. With the approaching of the O(6) limit the
ratio $\Delta E^{exp}_2$ decreases to values near 1.1 with
respective increase in the bandmixing interaction. For the region
of U(5)-O(6) [or E(5)] transition nuclei ($R^{exp}_4\sim 2.2$) and
nuclei near U(5) symmetry the analysis of our results indicate the
trend of sharply increasing interaction strength which may also be
considered as the hallmark of a completely rearranged structure of
collective spectrum.

In fig.~\ref{nN}  the values of the parameter of collectivity
``n'' obtained for Mo, Ru and Pd nuclei are plotted as functions
of the neutron number. It is clearly seen that from Mo to Pd
nuclei the respective curves are systematically shifted up,
demonstrating the overall move towards the vibrational collective
mode. In addition, in any particular group of isotopes we observe
a decrease in ``$n$'' with the increase of the neutron number $N$
towards the mid-shell regions. This result confirms the physical
significance of the quantity ``n'' as a characteristics of the
changing collectivity.

\section{Algebraic aspects of changing rotation/vibration collectivity}\label{On}

The proposed extended VBM application provides a test for the
change in nuclear dynamics between a rotation and vibration regime
of collectivity by assuming the presence of a ``universal''
interaction , $V'$, capable of accounting the relative
contribution of both modes. Hence the term $V'$ plays a crucial
role in the obtained systematics of nuclear collective properties,
which emphasizes the need for its deeper physical understanding.
This could be done by using the formalism of the Orthogonal group
$O(N)$ \cite{V88,CMV81} and its reduction to the rotational group
$O(3)$ in the chain\footnote{The notation ``$N$'' in this section
should not be mixed with the notation of the neutron number in the
other sections.}
\begin{equation}
O(N_1) \supset O(N_2) \supset ... \supset O(3) \ , \label{ONchain}
\end{equation}
where $N_1 > N_2 > ... > 3$.

From microscopic point of view the dimension $N$ of the group as
well as its subgroups in (\ref{ONchain}) might  be related to the
intrinsic configurations which contribute to the
coherent/vibrational behavior of the system. (For example these
could be configurations based on superfluid nucleon pairs that
give rise to vibrational phonon degrees of freedom.) In this
framework $O(N)$ can acquire the meaning of a rotation in the
$N$-dimensional space of intrinsic coordinates responsible for the
non adiabatic vibration component of the collective motion. (The
dimension of this space should increase towards nuclear vibration
regions).

From collective point of view, one can directly consider the
interaction $V'$ as a manifestation of nuclear boson degrees of
freedom. Then the different subgroups appearing in the chain
(\ref{ONchain}) should depend on the particular boson realization
and on the underlying collective dynamics of the system. As an
important consequence of the assumed bosonic character of the
interaction $V'$ only the totally symmetric irreducible
representations of $O(N)$ and its subgroups will play a role.
Therefore, the basis states for the chain (\ref{ONchain}) will be
characterized by the different single quantum numbers of the
different groups/subgroups (let us denote them by $\Lambda
_{N_i}$, with $i$ enumerating the different subgroups) and by the
possibly appearing additional quantum numbers accounting for
multipolarity problems in the reduction scheme.

In the boson framework the eigenvalues of the second order Casimir
operators, $\hat{C}_{N_i}$, for the different groups in
(\ref{ONchain}) will have the following form in the above basis
states
\begin{equation}
\langle \hat{C}_{N_i} \rangle=\Lambda _{N_i} (\Lambda _{N_i}
+{N_i}-2)\ , \label{Casim}
\end{equation}
with $i$ enumerating the different subgroups. Then we can
postulate that the Hamiltonian $\hat{V}'$ is a linear combination
of these operators
\begin{equation}
\hat{V}' = \sum_i A_{N_i} \hat{C}_{N_i} \ , \label{Onham}
\end{equation}

Further, following the reduction rules for the different quantum
numbers in the chain (\ref{ONchain}) and taking a particular set
of excited states (yrast, next to the yrast and so on) for
$\hat{V}'$, the quantum numbers $\Lambda _{N_i}$ can be determined
as linear functions of the angular momentum $L$ (which is the
quantum number of the lowest group, O(3), in the chain), namely
\begin{equation}
\Lambda _{N_i} =\alpha_{i} L+\beta_i. \label{Lin}
\end{equation}
The explicit form of the coefficients $\alpha_{i}$ and $\beta_i$
depends on the particular set of subgroups in (\ref{ONchain}) and
generally its derivation might be a complicated task.  After
introducing (\ref{Lin}) into (\ref{Casim}), the matrix elements of
$\hat{V}'$ in the considered states appear in the following
general form
\begin{equation}
<\hat{V}'>= \widetilde{A}L(L+\widetilde{B}) \ , \label{Onres}
\end{equation}
where $\widetilde{A}=\widetilde{A}(A_{N_i},\alpha_{i},\beta_i )$
and $\widetilde{B}=\widetilde{B}(A_{N_i},\alpha_{i},\beta_i )$ are
determined through the coefficients in (\ref{Onham}) and
(\ref{Lin}). Eq.~(\ref{Onres}) provides the same type of angular
momentum dependence as the one appearing in our modified VBM
energy expression. In this respect our model characteristics of
collectivity $n$ can be related to the quantity $\widetilde{B}$
appearing in (\ref{Onres}).

Here we illustrate the above scheme by considering the simplest
case of the group $O(5)$ which is a well known example appearing
in the context of the U(5) and  O(6) limits of IBM \cite{IA87,DB}.
So, we consider the reduction
\begin{equation}
O(5) \supset O(3) \supset O(2) \ , \label{O6chain}
\end{equation}
where the irreps of $O(5)$ and $O(3)$ are labelled by the quantum
numbers $\tau$ and $(L,M)$ respectively. Since $O(5)$ is not fully
decomposable to $O(3)$, an additional quantum number $\nu_\Delta$
is introduced. Thus the states associated to the chain
(\ref{O6chain}) are characterized by $|\tau \nu_\Delta LM>$ with
the reduction rule
\begin{eqnarray}
L&=& 2\lambda, 2\lambda-2, ..., \lambda +1, \lambda \\
\lambda &=& \tau - 3\nu _\Delta \ .
\end{eqnarray}
Then the Hamiltonian (\ref{Onham}) reads
\begin{equation}
\hat{V}'=A_{5}\hat{C}_{O(5)}+A_{3}\hat{C}_{O(3)} \ ,
\end{equation}
with
\begin{equation}
<\hat{V}'>=A_{5}\tau (\tau +3)+A_{3}L(L+1) \ . \label{O6IBM}
\end{equation}
By restricting ourselves with the yrast contributions of $V'$ we
have $\tau =L/2$ which yields
\begin{eqnarray}
<\hat{V}'> &=& {A_{5} \over 4}L(L+6)+A_{3}L(L+1) \nonumber  \\
          &=& \frac{1}{4}(A_5 +4A_3 )L\left(L+{6A_5 +4A_3 \over A_5
          +4A_3}\right)\ .
          \label{O5en}
\end{eqnarray}
Eq.~(\ref{O5en}) is in the form of eq.~(\ref{Onres}) with
$\widetilde{A}= (A_5 +4A_3)/4 $ and $\widetilde{B}=(6A_5
+4A_3)/(A_5+4A_3)$. If we assume that the inertial term
$\widetilde{A}$ is a constant then $\widetilde{B}$ can be
determined as a function of $A_5$
\begin{equation}
\widetilde{B}=\frac{5}{4\widetilde{A}}A_5 +1    \ . \label{Btilde}
\end{equation}
Eq.~(\ref{Btilde}) demonstrates in a simple way that for $A_5=0$
the quantity $\widetilde{B}$ reduces to 1, while for $A_5>0$ it
increases linearly.

From collective point of view the physical contents of the above
result could be sought in relation to the 5-dimensional harmonic
oscillator which is the natural framework in the study of nuclear
quadrupole vibrations. In such a way the increasing contribution
of the $O(5)$ term in the interaction $V'$ could be directly
interpreted as the increasing influence of vibrational degrees of
freedom in the collective motion of the system.

As has been mentioned above, the deeper structure interpretation
of the nonadiabatic term $V'$ might require the involvement of
larger-dimensional orthogonal groups and therefore more
complicated treatment of the underlying formalism.

The above analysis gives at least an intuitive idea about the ways
in which the non-rotational degrees of freedom can be involved in
the consideration away from the typical SU(3) rotational region.
It suggests a relevant tool in the study of the changing
collective properties of nuclei. As a most general approach in
this direction it might be valuable to consider a model space
containing both O(N) and SU(3) spaces. It could allow a reasonable
treatment of the total Hamiltonian wave functions with subsequent
model predictions for the electromagnetic transitions. Even now we
can outline qualitatively the main characteristics of their
behavior. So, as far as $V'$ is diagonal with respect to the SU(3)
basis states its input will affect mainly the behavior of
intraband transitions in gsb and the $\gamma$-band, which might be
expected to deviate from a rotation-like angular momentum
dependence towards vibrational. In this scheme the interband
transitions would be only indirectly affected through the band
mixing interaction following the assumption that the same
gsb-$\gamma$ band coupling mechanism is conserved with the change
of collectivity. Although in the considered nuclear regions there
is not yet enough experimental information on these transitions,
such a development would be of use in the understanding the
mechanism of the changing nuclear collectivity at all.

\section{Conclusion}\label{co}
In the present paper we propose an extended model formalism for
description of the low-lying ground and $\gamma$-band collective
states and their band mixing interactions in a wide range of
nuclei starting from the SU(3) rotation regions and reaching
nuclei near the almost vibrational region.

On this basis we obtain model description and implement detailed
analysis of the ground -- $\gamma$-band structure in the region of
Mo, Ru and Pd isotopes. As a result we reproduce accurately not
only the levels up to the back-bending region but also the
respective experimentally observed odd-even staggering effect in
the gamma bands.

The analysis of the results obtained allows us to draw the
following main conclusions.

In the considered  region ${40 <}$Z${< 50}$ the $\gamma$ -- gsb
interaction strength increases away from the SU(3) symmetry with
the approaching of the transitional and vibrational collective
regions. It conserves the same band coupling mechanism typical for
the VBM scheme although in this region the overall structure of
the spectrum is changed in consistence with the multiplet
characteristics inherent for the O(6) and U(5) collective nuclei.
The systematic analysis of the mutual ground $\gamma$-band
disposition (characterized by the quantity $\Delta E_2$) provides
detailed quantitative characteristics of the  different regions of
collectivity in Pd, Ru and Mo isotopes. The respective changes in
the structure of the spectrum reflect in the respective fine
odd-even staggering patterns observed there.

The implemented theoretical analysis outlines the evolution of
collectivity along the considered isotopic chains. The obtained
values of the model characteristics $n$ provide information about
the way of deviation from the SU(3) symmetry. The increase in $n$
is correlated with the systematic behavior of the gsb ratio of
collectivity $R_4$ indicating a rapid decrease in $R^{exp}_4$
betwen $3.33 \geq R^{exp}_4 > 2.5$, saturation of $R^{exp}_4$ near
$2.5$ and signs of further slow decrease towards $R_4\leq 2.2$.
Thus it provides a specific map of collectivity for the ground --
$\gamma$-band structure covering the  regions with nearly-SU(3)
nuclei, U(5)-SU(3) [or X(5)] and O(6) symmetry nuclei, and some
nuclei towards U(5)-O(6) [or E(5)] and nearly-U(5) symmetry
regions.

The algebraic framework of the study allows further development
with a detailed treatment of the collective and intrinsic degrees
of freedom responsible for changing of nuclear collectivity from
rotation to the vibration mode. In this respect the use of the
formalism of orthogonal groups seems to be promising. Such an
approach could provide a relevant tool for further analysis of
available and newly obtained experimental data and their
interpretation in terms of the above exact and transition
symmetries.

\section{Acknowledgments}
The authors are grateful to S. Drenska for the stimulating
discussions at various stages in the development of this work.
Also, the authors are thankful to D. Bonatsos for the useful
discussions and the careful reading of manuscript.

\end{document}